# COMPARISONS OF EQUIVALENT CIRCUIT PREDICTIONS WITH MEASUREMENTS FOR SHORT STACKS OF RDDS1 DISCS, AND THEIR POTENTIAL APPLICATION TO IMPROVED WAKEFIELD PREDICTION

R.M. Jones[1], SLAC; T. Higo, Y. Higashi, N. Toge, KEK;
N.M. Kroll[2], SLAC & UCSD; R.J. Loewen[1], R.H. Miller[1], and J.W. Wang[1]; SLAC

*Abstract*

In fabricating the first X-Band RDDS (Rounded Damped Detuned Structure) accelerator structure, microwave measurements are made on short groups of discs prior to bonding the discs of the entire structure. The design dispersion curves are compared with the frequency measurements. The theory utilised is based on a circuit model adapted to a short stack of slowly varying non-uniform discs. The model reveals the nature of the modes in the structure and may also be used to refit the experimental data to the parameters in a model of the wakefield given earlier [1]. This method allows a more faithful determination of the wakefield that a beam will experience as it traverses the structure. Results obtained on the frequencies are compared to the original design.

## 1. INTRODUCTION

The design and fabrication of the first RDDS accelerator structure (RDDS1) is described in [2]. The basic fabrication units of the structure are discs consisting of a rounded beam iris and a pair of identical half cells one on each side of the iris. The disc also includes the four circular waveguide sections which form the four manifolds together with the coupling slots which connect them to the cells. Further details are found in [2] and [3]. Microwave measurements were performed during the fabrication procedure to provide quality assurance. These were of two sorts. The first consisted of resonance frequency measurements on a single disc terminated with flat conducting plates in pressure contact with the disc [3]. The second consisted of similar measurements on stacks of six successive discs (eg disc n to disc n+5) carefully aligned and pressed together between two flat conducting plates. Each end plate was provided with an off center probe, similarly placed within the cell (rather than the manifold) region, and at an azimuth such that they faced one another. Both monopole and dipole resonant frequencies were determined by means of of network analyser measurements of scattering matrix parameters, (primarily $S_{12}$) between the end plate probes.

The single disc measurements provide resonant frequencies for the zero and $\pi$ modes of the lowest monopole band, the $\pi$ mode of the first dipole band, and the zero mode of the second dipole band. The primary information that one might hope to aquire from the single disc measurements would be the accelerating mode frequency (ie the $2\pi/3$ monopole), which is supposed to be the same for each disc, and the tuning profile of the synchronous frequency of the lowest dipole mode. Although this information is not provided directly, the accelerating mode frequency is expected to be given to adequate accuracy from the monopole mode frequencies $f_0$ and $f_\pi$ by the formula

$$f_{acc} = 2f_0 f_\pi / \left(3f_0^2 + f_\pi^2\right)^{1/2} \qquad (1.1)$$

The lower dipole pi mode frequency is very close to the synchronous frequency, so that its profile can provide an adequate surrogate for the synchronous mode profile. Also, these two frequencies uniquely determine two of the equivalent circuit parameters (independently of the values of the other parameters), which may then be compared to profiles for these two parameters obtained by interpolation formulas from the five cells actually simulated. Further details regarding the application of the single disc measurements may be found in [3], especially on the development of a rapid routine quality assurance procedure that could be integrated into a manufacturing process.

With the six disc stacks, more resonances can be observed, thereby providing more complete information regarding the accuracy of machining, simulation, and equivalent circuit representation. Of particular interest is the fact that each of the stacks should, despite the fact that they are made up of discs dimensioned to fit the detuning profile, should have a $2\pi/3$ phase advance monopole mode all with the same 11.424 GHz frequency of the acceleration mode. Since the completed structure is driven at 11.424GHz, a frequency error in the acceleration mode of a particular stack will translate into a phase advance error. Because these measurements were performed while fabrication was in progress, compensating dimensional changes could be made in subsequently fabricated cells so as to prevent the accumulation of phase errors [2]. The stacks also provide more detailed information about the Brillouin diagram of the dipole modes. As discussed below, this information will be used to obtain more accurate information on the synchronous frequency profile and on the coupling between the lower dipole modes and the manifold.

---

[1] Supported under U.S. DOE contractDE-AC03-76SF00515.
[2] Supported under U.S. DOE grant DE-FG03-93ER407.

## 2. FEATURES OF N DISC STACKS

We first consider the case of a uniform N disc stack. Our attention will be focussed on the modes associated with the first monopole band, the first two dipole bands, and the first bands associated with the four manifolds. These are the low lying modes found in single disc simulations with periodic boundary conditions, their bands being traced out by varying the phase advance. (Historically, one quarter of a single cell is simulated, with symmetry-symmetry boundary conditions for the monopole and a pair of nearly degenerate manifold bands, and metallic-symmetry boundary conditions for the lower two dipole bands and a manifold band. The results would be identical for a disc.) The modes of an N disc stack terminated at each end by a conducting plate are equal amplitude superpositions of the two oppositely directed travelling wave solutions. One may think of one boundary as determining their relative phase, and then the relative phase required by the other boundary determines which phase advances correspond to modes. The "band edge" solutions (ie 0 and $\pi$ phase advance) are standing rather than running waves, and are missing or included in the mode spectrum accordingly as they do or do not fit the imposed boundaries. For the N disc stack one expects modes corresponding to phase advances $(n/N)\pi$, with n = 0,1...N for the monopole band modes, n = 1,...N-1 for the modes of the manifold bands. The differences arise from the fact that the former are TM and the latter TE. The dipole bands are TE-TM hybrids, but at the band edges one or the other predominates. The consequence is that n = 0 is missing from the lower dipole band and n = pi is missing from the upper dipole band. The four modes of the N = 1 case discussed above are illustrative of these rules. The total number of modes of an N disc stack associated with the bands specified is 9N - 3 (counting both dipole orientations).

Experimental study of a uniform stack would provide a check on the Brillouin diagram obtained from simulation. Also for N a multiple of three the sequence of phase advances represented includes $2\pi/3$. Thus one gets a better check on the accelerating mode frequency by studying such stacks. The RDDS1 disc fabrication was checked for the entire structure by measuring 34 six disc stacks. The measured stacks belonged to the sequence of detuned cells and were therefore not uniform, but as mentioned above this does not affect the frequency of the accelerating mode. For the monopole band observations, only the accelerating mode frequency was recorded. The frequencies of all dipole modes which could be seen from the probes were also recorded. These included the six lower and upper dipole modes with probe coupled orientation, and those dipole phased manifold modes strongly enough coupled to the cells to be seen from the probes

## 3. EQUIVALENT CIRCUIT ANALYSIS OF DETUNED N DISC STACKS.

The equivalent circuit of [1] was designed to represent the first two dipole bands, and the first manifold band of the DDS and RDDS structures. (Because the manifolds include a transmission line in their representation, there are solutions to the circuit equations which refer to higher order manifold bands, but they have not been tailored to represent the actual higher order manifold bands with any reliability.) The actual structure contains a degenerate pair for each of these bands, but the circuit models only one of them. The circuit for an N = 3 disc stack is illustrated in Fig. (1). Here, to conform to earlier notations, we number the discs 1/2,..N-1/2. The full LC circuits between discs n-1/2 and n+1/2 (n =1,...,N-1) represent the hybrid TE-TM modes of the cells between the discs. Their loop currents are represented by the amplitudes $a_n$, $\hat{a}_n$ respectively for the TE and TM circuits respectively. The shunted transmission line sections n correspond to the portion of the manifold adjacent to the n'th cell and are represented by the amplitude variables $A_n$ proportional to the voltages across the shunt. The half cells at the ends of the TM chain with doubled C and halved L correspond to the region between the end discs and the shorting plates, with amplitudes represented by $\hat{a}_0$ and $\hat{a}_N$, and the manifold transmission lines are shorted at a distance one period away from the adjacent shunts.

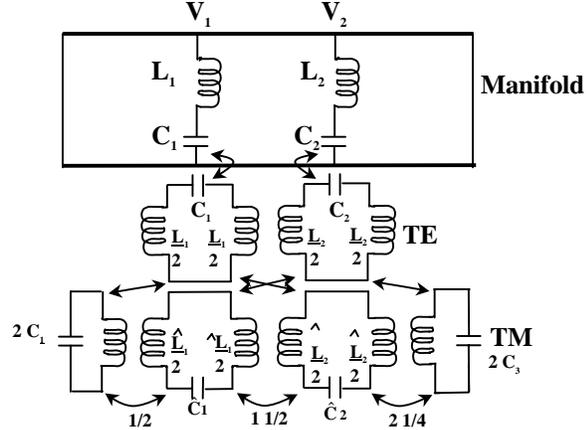

Figure 1: Circuit diagram of 3-cell stack

Analogous to [1] the homogeneous circuit equations may be written in the form:

$$RA - Ga = 0 \qquad (1.2)$$
$$(H - f^{-2})a + H_x \hat{a} - GA = 0 \qquad (1.3)$$
$$(\hat{H}' - f^{-2})\hat{a} + \hat{H}'^{t}_x a = 0 \qquad (1.4)$$

Here A, a, and $\hat{a}$ are N-1, N-1, N+1 component column vectors respectively. R, G, and H are N-1 x N-1 matrices, $\hat{H}'$ is N+1 x N+1, while the matrices $H_x$ and $\hat{H}'^{t}_x$ are N-1 x N+1 and N+1 x N-1 respectively. The matrix elements of these matrices are the same as those given in [1] except that the primes on $\hat{H}$ and $\hat{H}_x^{t'}$ indicate that their (0 1) and (N+1 N) matrix elements are doubled. This asymmetry in the equations could be removed by rescaling $\hat{a}_0$ and $\hat{a}_N$, but the form of the eigenvectors is simpler with the equations as they are. In particular, in the case of a uniform stack direct substitution in (2), (3), and (4) verifies that the eigenvectors take the form

$a_n = K_1 \cos(n\psi);$   $n = 0,..N$   (1.5)
$a_n = K_2\ A_n = \sin(n\psi);$   $n = 1,...,N-1$   (1.6)
with $\psi = (m/N)\pi;$   $m = 0,...,N.$   (1.7)

The modal frequencies are determined by the phase (psi)-frequency dispersion relation of Eq. (10) in [1]. As indicated there, its three lowest roots provide the Brillouin diagram of the lower two dipole bands and the lowest manifold band. An example appears later as Fig. (2).

Here the modal frequencies are determined by substituting the $\psi$ values of Eq. (7). The specification of the eigenvectors is completed by computing $K_2$ from Eqs. (2) and (6) and then $K_1$ from (5) and (4). For two of the three roots at $\psi = 0$ and at $\psi = \pi$, $K_1 = 0$. All amplitude variables then vanish, and hence

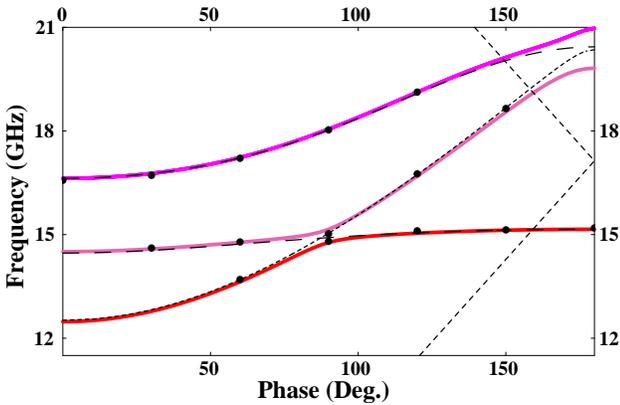

Figure 2: Brillouin diagram corresponding to RDDS1 cell stack 98 to 103 (average cell 100.5). The points are obtained from an experimental measurement and the lines are obtained from the circuit model in which the original design was prescribed prior to the experiment.

these frequencies do not represent modes of the stack The remaining frequencies at these phase values represent pure TM modes for which $K_1$ can be assigned an arbitrary non-zero value. For the detuned case numerical methods must be used to determine both the eigenvectors and the modal frequencies. Equations (2, 3, and 4) are a set of 3N-1 linear homogeneous equations in the 3N-1 amplitudes represented by a, $\hat{a}$, and A. Modal frequencies were determined by finding the frequencies at which the determinant of the coefficients vanishes. This root search process was greatly facilitated by starting from the modal frequencies of a uniform stack with parameters corresponding to those of the average cell in the stack. These may be obtained from the interpolation procedure described in [4]. Once the modal frequencies are known determination of the eigenvectors is a well known and numericlly efficient linear algebraic procedure. The shift of the modal frequencies from those of the average cell uniform stack is usually quite small, in which case the individual modes can still be designated by band and phase advance. Ambiguities can arise when there is near degeneracy. Although distorted by detuning, equivalent circuit eigenvectors can help resolve them. Likewise, the phase and magnitude of S_12 at the resonant peaks can do the same for the stack experimental measurements

We conclude with an example based on the stack formed by the six discs 98 to 103. The space between disc 100 and 101 constitutes the average cell, which we designate as cell 100.5. Its Brillouin diagram is shown in Fig. (2), and experimental points are also plotted. Table 1 provides a numerical comparison between experimental points, detuned stack equivalent circuit computed points, and points for a uniform 6 disc stack with parameters for cell 100.5. The particular frequencies selected for display in the table are thought to be the most relevant for wake prediction because of their bearing on cell to manifold coupling and synchronous frequency identification. The equivalent circuit computed shifts of the detuned stack frequencies from the uniform average stack frequencies are indeed quite small in this case, so that the association of the detuned stack frequencies with the uniform stack phas advances and bands seems quite unambiguous. Also the eigenvectors do provide support for the identification, although some of the patterns do show considerable distortion. The identification of the experimental frequencies with phase advances and bands is so far based primarily on the pattern. The association of the lower 60 degree and higher 120 degree frequencies with the manifold is, however, supported by the small amplitude of their $S_{12}$ peaks. We note that associated with each stack we will have typically 27 measured frequencies, 12 from the single disc and usually 15 from the six cell stack. It is our intention to use this data to refine our parameter interpolation curves and our synchronous frequency profile, and to thereby improve our wake predictions, but we do not yet have a tested methodology for doing so.

| $\psi$ | 60 | 90 | 120 | 150 | 180 |
|---|---|---|---|---|---|
| $f_{exp}$ | 13.7 14.7856 | 14.8048 15.0292 | 15.110 16.759 | 15.1388 | 15.1923 |
| $f_{mode}$ | 13.6358 14.7686 | 14.7803 15.1544 | 15.0399 16.7701 | 15.1256 | 15.1556 |
| $f_{av}$ | 13.6356 14.7759 | 14.7759 15.1539 | 15.0404 16.6918 | 15.1271 | 15.1537 |

Table 1: Experimentally measured stack frequencies, $f_{exp}$, modally determined frequencies, $f_{mode}$ and, average cell frequencies.